\documentclass{article} 
 
\usepackage{amsmath,amssymb,amsthm,latexsym} 
 
\usepackage{stmaryrd,wasysym,upgreek,mathrsfs,dsfont} 
\usepackage[english]{babel} 
\usepackage{graphicx,color} 
\newcommand{\R}{\mathbb{R}} 
 
\usepackage[pdftex]{hyperref} 

\newtheorem{lemma}{Lemma}[section]
\newtheorem{theorem}{Theorem}[section]
\newcommand{\beq}{\begin{equation}}
\newcommand{\eeq}{\end{equation}}
\newcommand{\beqa}{\begin{eqnarray}}
\newcommand{\eeqa}{\end{eqnarray}}

\newcommand{\si}{\sigma}

\newcommand{\Tr}{{\rm Tr}}

\begin{document} 
\title{Constructive Matrix Theory} 
\author{V. Rivasseau\\
Laboratoire de Physique Th\'eorique, CNRS UMR 8627\\ 
Universit\'e Paris XI,  F-91405 Orsay Cedex, France}
 
\maketitle 
\begin{abstract} 
We extend the technique of constructive expansions
to compute the connected functions of matrix models in a uniform way as the size
of the matrix increases. This provides the main missing ingredient
for a non-perturbative construction of the $\phi^{\star 4}_4$ field
theory on the Moyal four dimensional space.
\end{abstract} 
 
\section{Introduction} 

Constructive field theory build functions whose Taylor expansion 
is perturbative field theory \cite{GJ,Riv1}. Any formal power series
being asymptotic to infinitely many smooth functions, perturbative field theory alone
does not provide any well defined mathematical recipe to compute to arbitrary accuracy
any physical quantity, so in a deep sense it is no theory at all.

In field theory infinite volume quantities
are expressed by connected functions. One main advantage of perturbative field theory 
is that connected functions are simply the sum of the connected Feynman graphs.
But the expansion diverges because there are too many such graphs.

In fact connectedness does not require the full knowledge of a Feynman graph 
(with all its loop structure) but only the (classical) notion of a spanning tree in it. 
To summarize constructive theory, let's say that it is all about 
working as much as possible with the trees only, and resumming or hiding 
most of the quantum loops. This is the constructive golden rule:

\emph{``Thou shall not know all the loops, or thou shall diverge!"}

However the constructive program launched by A. Wightman and pursued by J. Glimm, A. Jaffe
and followers in the 70's was a partial failure because no natural four dimensional field theory 
could be identified and fully built. This is because only non-Abelian gauge theories 
are asymptotically free in the ultraviolet limit. But ultraviolet asymptotic freedom also means
infrared slavery, and non-perturbative long range effects such as quark confinement 
are not fully understood until now, even at a non-rigorous level. 
The constructive program went on, but mostly as a set of rigorous techniques applied to many 
different areas of mathematical physics \cite{constr1,constr2}. 

Recently quantum field theory on non-commutative space has been shown renormalizable.
The simplest such theory is the $\phi^4_4$ theory on the Moyal space, 
hereafter called $\phi^{\star 4}_4$. Grosse and Wulkenhaar  \cite{GrWu1} overcame the main 
obstacle to renormalizability, namely the ultraviolet/infrared mixing, 
through the use of a new symmetry called Langmann-Szabo duality \cite{LaSz}. 
Following their initial breakthrough, a flurry of papers
has appeared to extend this result to other models and to generalize
to the Moyal context many useful tools and techniques
of ordinary perturbative field theory. For recent reviews, see \cite{RV,Riv2}.

It now appears that four dimensional non commutative field theories
are \emph{better} behaved than their commutative counterparts. 
In particular $\phi^{\star4}_4$, in contrast to
its commutative counterpart, is \emph{asymptotically safe} \cite{GrWubeta,DR1,DGMR}:
the flow between the bare and the renormalized
coupling constant is bounded. In fact the graphs responsible for the flow of the coupling
constant compensate exactly at any order with those responsible for 
the wave function renormalization. This is an exciting discovery:
LS symmetry may play a role similar to supersymmetry in taming ultraviolet flows.

Asymptotic safeness is in a sense much simpler than asymptotic freedom, and
$\phi^{\star4}_4$ now stands out as an obvious candidate for a four 
dimensional constructive field theory without unnatural cutoffs 
(although on the unexpected Moyal space).

But after \cite{DGMR} one main difficulty remained unsolved 
on the road to constructive $\phi^{\star4}_4$. 
Current cluster expansions used in standard bosonic constructive theory \cite{Riv1} 
are unsuited to treat matrix models with large number of components. To explain why,
let us compare the large $N$ vector $\phi^4$ model and the large $N$ matrix\cite{Hoo} $\phi^4$ model. In both cases the coupling scales as $1/N$  for a non trivial limit as $N$ gets large:
at order $n$ in a graph there is indeed in both cases at most about $n$ loops of indices.
But in the first case the field has $N$ vector components, and at a given vertex
only two different colors can meet. Knowing only a spanning tree in the graph, it is 
still possible to sum over all indices at the right cost. To do this, at any
leaf of the tree one can sum over the index which does not go 
towards the root and keep the other one for the next step. 
Iterating from leaves of the tree towards the root there is only one index summed per vertex, 
(except at the root, where in the case of a vacuum graph there are two indices to sum, leading
to the final global $N$ factor of vacuum graphs).
This procedure does not violate the constructive golden rule, as no loops need to be known.

But a matrix model is very different. The field has $N^2$ components 
and at a given vertex four different indices meet. The scaling of the vertex is still only 
$1/N$, but this is because each propagator identifies \emph{two} matrix field indices with two others,
rather than one. Therefore matrix models apparently clash 
with the constructive golden rule. The knowledge of the full loop structure 
of the graph, not only of a tree, seemed until now necessary to recover the correct power counting,
for instance a single global $N^2$ factor for vacuum graphs.

Since $\phi^{\star 4}_4$ is a quasi-matrix model with a large number of components 
in the ultraviolet limit \cite{GrWu2} it is plagued with this constructive matrix difficulty,
hence seems unsuited at first sight for a constructive analysis.
The difficulty persists in the direct space version \cite{GMRV} of the model, but
in a different guise. In that representation, it is the non-locality
of the vertex in $x$ space which is impossible to treat with standard constructive
methods, such as ordinary cluster and Mayer expansions with respect to lattices of cubes.

In short a new kind of expansion based on a new idea is required for constructive 
$\phi^{\star 4}_4$. This is what we provide in this paper.

The idea is in fact quite simple. Matrix models can be decomposed with
respect to an intermediate matrix field. Integrating over the initial field leads
in a standard way to a perfect gas of so called \emph{loop vertices} for this intermediate 
field. One can then perform the tree expansion directly on these loop vertices.
All indices loops then appear as the correct number 
of traces of products of interpolated resolvents, which can be bounded 
because of the anti-Hermitian character of the intermediate field insertions.

We take as an example the construction of the connected functions
of a matrix model perturbed by a $\frac{\lambda}{N} \Tr \phi^\star \phi \phi^\star \phi$ interaction. We prove 
as a typical result Borel summability in $\lambda$ of the normalization and 
of the connected $2p$ point functions
\emph{uniformly in the size of the matrix}
\footnote{Non-uniform Borel summability, taking $\lambda$
smaller and smaller as $N \to \infty$ is trivial and would completely 
miss the difficulty.}.

In a companion paper \cite{MR} we explore the consequences of this idea in
the more traditional context of commutative constructive field theory.

Recall that it is possible to rearrange Fermionic perturbation theory 
in a convergent expansion \emph{order by order} by grouping together
pieces of Feynman graphs which share a common tree \cite{Les,AR2}.
But bosonic constructive theory cannot be simply rearranged in such a convergent way 
\emph{order by order}, because all graphs at a given order
have the same sign. Resummation of the perturbation
theory (which occurs only e.g. in the Borel sense) 
must take place between infinite families of graphs (or subparts of graphs)
of different orders. To explicitly identify these families seemed until now almost impossible. 
Cluster and Mayer expansions perform this task but in a very 
complicated and indirect way, through an intermediate discretization of
space into a lattice of cubes which seems \emph{ad hoc} for 
what is after all a rotation invariant problem.

In fact the cluster expansion between loop vertices, although 
found in the context of matrix models, can identify such families also in the ordinary commutative case
\cite{MR}. This simplifies traditional bosonic constructive
theory, avoiding any need for cluster and Mayer expansions.
We should bring in this way Bosonic constructions
almost to the same level of simplicity than the Fermionic ones and explore
the consequences in future publications.

\section{Matrix Model with Quartic Interaction}

The simplest $\phi^4$ matrix model is 
a Gaussian independent identically distributed measure on $N$ by $N$
real or complex matrices perturbed by a positive
$\frac{\lambda}{N} \Tr \phi^\star \phi \phi^\star \phi$
interaction. The $N \to \infty$ limit is given by planar graphs. It can be studied through various 
methods such as orthogonal polynomials \cite{Meh,Voi}, supersymmetric 
saddle point analysis \cite{Efe,Mir,Zuk} and so on.
However none of these methods seems exactly suited to 
constructive results such as Theorem \ref{borelunif} below.

Consider the complex case (the real case being similar). The normalized interacting measure is
\begin{equation} \label{functional}
d\nu (\Phi) = \frac {1}{Z(\lambda,N)}
e^{-\frac {\lambda}{N} 
\Tr \Phi^\star \Phi \Phi^\star \Phi} d\mu(\Phi) 
\end{equation}
where 
\begin{equation} d\mu = \pi^{-N^2}
e^{-\frac 12 \Tr \Phi^{\star}  \Phi } \prod_{i,j} d \Re \Phi_{ij} d \Im \Phi_{ij} 
\end{equation}
is the normalized Gaussian measure with covariance
\begin{equation}
<\Phi_{ij} \Phi_{kl}>= <\bar \Phi_{ij} \bar \Phi_{kl}>  = 0, \ \ 
<\bar \Phi_{ij} \Phi_{kl}> = \delta _{ik} \delta_{jl} .
\end{equation}

For the moment assume the coupling $\lambda$ to be real positive and small. 
We decompose the $\Phi$ functional integral according to an intermediate
Hermitian field $\sigma$ acting either on the right or on the left index. 
For instance the normalization $Z(\lambda, N)$ can be written as:
\begin{equation}
Z(\lambda, N)  = \int d\mu_{GUE}(\si^R) 
e^{- \Tr \log (1\otimes 1 + i \sqrt{\frac{\lambda}{N}} 1 \otimes \sigma^R ) }
\end{equation}
where $d\mu_{GUE}$ is the standard Gaussian measure on an Hermitian
field $\sigma^R$, that is the measure with covariance 
$<\sigma^R_{ij} \sigma^R_{kl}>= \delta _{il} \delta_{jk}$. The $e^{-\Tr\log }$ 
represents the Gaussian integration over $\Phi$, hence a big $N^2$ by $N^2$
determinant. It is convenient to view  $\mathbb{R}^{N^2}$ 
as $\mathbb{R}^{N}\otimes \mathbb{R}^{N}$.
For instance the operator $H=\sqrt{\frac{\lambda}{N}}[1 \otimes \sigma^R ]   $
transforms the vector $e_{m}\otimes e_{n}$ into $\sqrt{\frac{\lambda}{N}}e_{m} \otimes \sum_k \sigma^R_{kn} e_k$.
Remark that this is an Hermitian operator because $\sigma^R$ is Hermitian.

By duality of the matrix vertex, there is an exactly similar formula 
but with a left Hermitian field $\sigma^L$ acting on the left index, 
and with  $[\sigma^L \otimes 1 ]$ replacing $ [1 \otimes \sigma^R ]$.
From now on we work only with the right field and drop the $R$ superscript for simplicity.

We want to compute e.g. the normalization $Z(\lambda, N)$, which is the (Borel) sum
of all connected vacuum graphs.
We define the  loop vertex $V$  by
\begin{equation}\label{vertex}
V=- \Tr \log (1\otimes 1 + 1 \otimes i H ) ,
\end{equation}
and expand the exponential as $\sum_n \frac{V^n}{n!} $. To compute the connected
graphs we give a (fictitious) index $v=1,..., n$ to all the $\sigma$ fields of a given 
loop vertex $V_v$. At any order $n$ the functional integral over $d\nu (\sigma)$ 
is obviously also equal to the same integral but with a Gaussian measure $d\nu (\{\sigma^v\})$ with degenerate covariance $<\sigma^v_{ij} \sigma^{v'}_{kl}>= \delta _{il} \delta_{jk}$. 
We apply then the \emph{forest} formula of \cite{AR1} to test connexity between the 
loop vertices from 1 to $n$. The logarithm of the partition function or pressure is then given
by the corresponding \emph{tree} formula exactly like in the Fermionic case \cite{AR2}.

\begin{theorem}
\begin{eqnarray}\label{treeformul}
\log Z(\lambda, N) = 
\sum_{n=1}^{\infty} \sum_T \bigg\{ \prod_{\ell\in T}   
\big[ \int_0^1 dw_\ell \sum_{i_\ell, j_\ell, k_\ell, l_\ell} \big]\bigg\} 
\int  d\nu_T (\{\sigma^v\}, \{ w \})  \nonumber \\
 \bigg\{ \prod_{\ell\in T} \big[ \delta _{i_\ell l_\ell} \delta_{j_\ell k_\ell}
 \frac{\delta}{\delta \sigma^{v(\ell)}_{i_\ell, j_\ell}}
 \frac{\delta}{\delta \sigma^{v'(\ell)}_{k_\ell, l_\ell}} 
\big] \bigg\} \prod_v V_v 
\end{eqnarray}
where 
\begin{itemize}

\item each line $\ell$ of the tree joins two different loop
vertices $V^{v(\ell)}$ and $V^{v'(\ell)}$,

\item the sum is over trees over $n$ vertices, which have therefore
$n-1$ lines,

\item the normalized Gaussian 
measure $d\nu_T (\{\sigma_v\}, \{ w \})  $ over the vector field $\sigma_v$ has covariance 
$$<\sigma^v_{ij} \sigma^{v'}_{kl}>= \delta _{il} \delta_{jk}w^T (v, v', \{ w\})$$
where $w^T (v, v', \{ w\})$ is 1 if $v=v'$,
and the infimum of the $w_\ell$ for $\ell$ running over the unique path from $v$ to $v'$ in $T$
if $v\ne v'$. This measure is well-defined because the matrix $w^T$ is positive.

\end{itemize}
\end{theorem}

This is indeed the outcome of the tree formula of \cite{AR1} in this case. 
This formula is convergent for $\lambda$ small enough!

\begin{theorem}\label{maintheor}
The series (\ref{treeformul}) is absolutely convergent for $\lambda$ small enough.
\end{theorem}

\begin{figure}[!htb]
\centering
\includegraphics[scale=0.25,angle=-90]{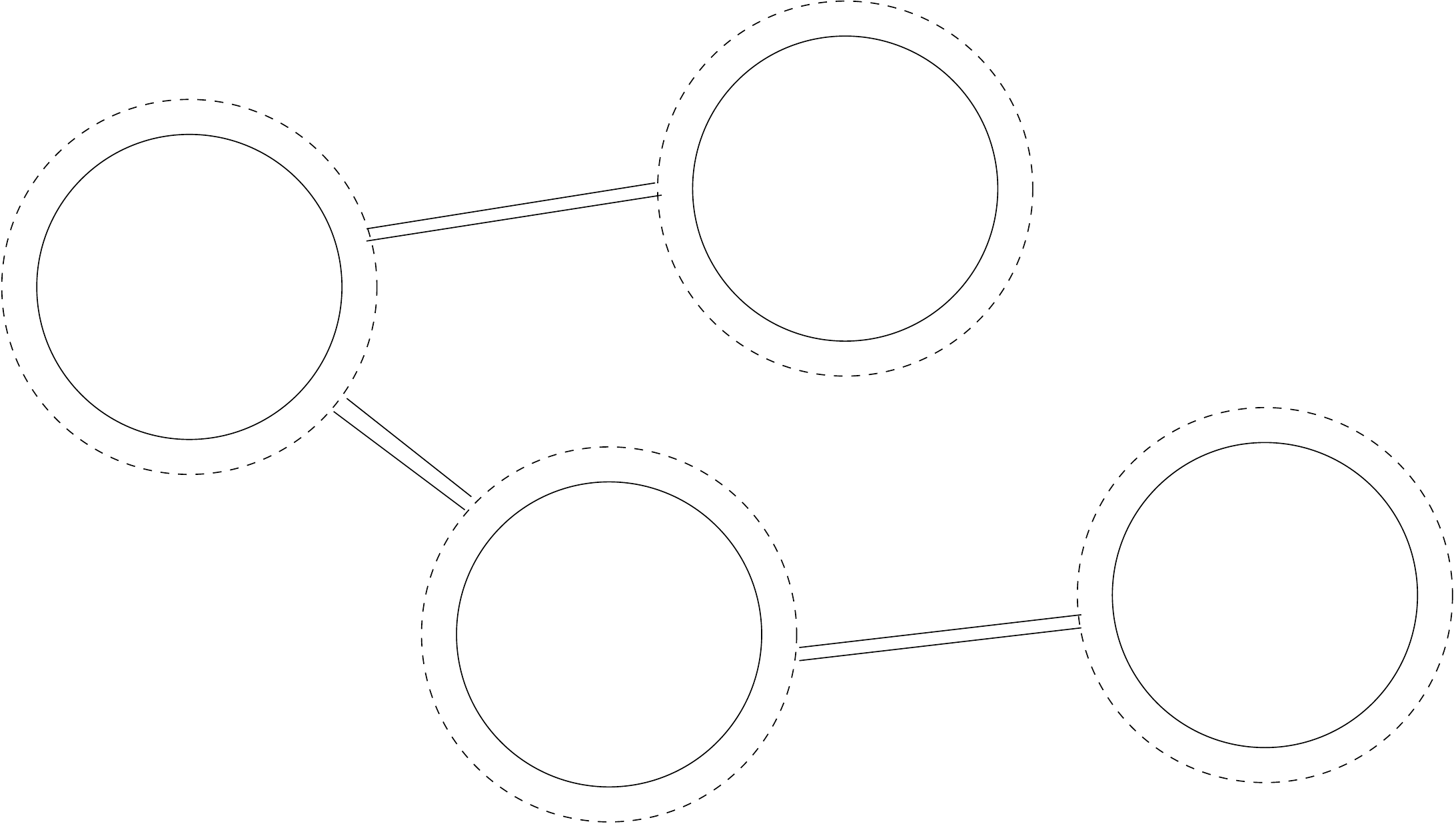}
\caption{A tree on four loop vertices}
\label{tree}
\end{figure}

\noindent{\bf Proof}\ 
Consider a vertex $V_v$ of coordination $k_v$ in the tree. Because the $\sigma $
field acts only on right indices, and left indices are conserved, there is a single
global $N$ factor for $V_v$ coming from the trace over the left index. We can then 
from now on essentially
forget about the left indices  except that they give a particular cyclic order
on $V_v$. See Figure \ref{tree} for a tree on four loop vertices, hence with three lines.

We compute now the outcome of the $k_v$ derivatives 
$\prod_{i=1}^{k_v}\frac{\delta}{\delta \sigma^{i}}$ acting on $V= -\Tr\log (1+iH)$
which created this vertex. Fix an arbitrary root line $\ell_0$ in the tree $T$.
There is a unique position $i=1$ on the loop vertex
from which a path in $T$ goes to $\ell_o$, and the loop vertex factor $V_v$ 
after action of the derivatives is 
\begin{equation}\label{loopvertex}
[\prod_{i=1}^{k_v}\frac{\delta}{\delta \sigma^{i}} ]    V_v =
N (-i \sqrt {\lambda /N})^{k_v}  \prod_{i=1}^{k_v}  C(i,i+1; \sigma^v)  
\end{equation}
where the cyclic convention is $k_v+1 =1$, and 
the operator $C(i, i+1; \sigma^v) = (1+iH(\sigma^v))^{-1}(j_i, j_{i+1})$ acts only on the right index
(it is no longer a tensor product, since the left trace has been taken into account in the 
global $N$ factor in front of $V_v$).

\begin{figure}[!htb]
\centering
\includegraphics[scale=0.3,angle=-90]{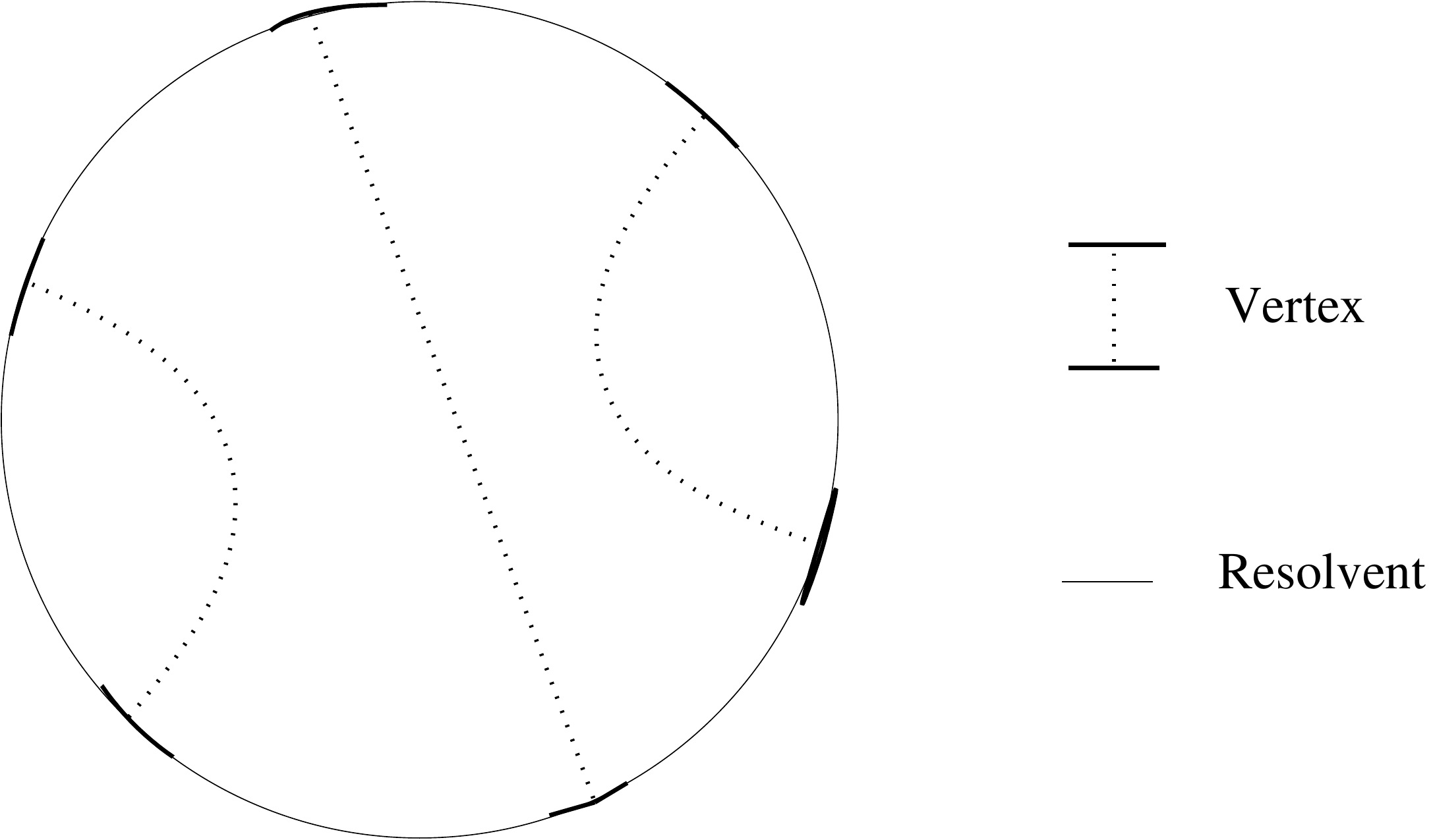}
\caption{Turning around a tree with four vertices and three lines}
\label{turnaround}
\end{figure}

To bound the integrals over all sums $\sum_{i_\ell, j_\ell, k_\ell, l_\ell}$ (which by the way
are only about right indices) we need now only a very simple lemma:

\begin{lemma}\label{key}
For any $\{w\} $ and $\{\sigma^v \}$ we have the uniform bound  
\begin{equation}
\vert \prod_{\ell \in T} \sum_{i_\ell, j_\ell, k_\ell, l_\ell}
\big[ \delta _{i_\ell l_\ell} \delta_{j_\ell k_\ell}
 \frac{\delta}{\delta \sigma^{v(\ell)}_{i_\ell, j_\ell}}
 \frac{\delta}{\delta \sigma^{v'(\ell)}_{k_\ell, l_\ell}} 
\big] \bigg\} \prod_v V_v   \vert  \le N^2
\end{equation}
\end{lemma}
\noindent{\bf Proof}\ 
Since $iH$ is anti-hermitian we have indeed $\Vert (1+iH)^{-1}\Vert \le 1 $.
The product over all vertices of the resolvents $C(i,i+1; \sigma^v)$ together with all
the sums $\sum_{i_\ell, j_\ell, k_\ell, l_\ell}$
exactly forms a big trace of $2(n-1)$ operators 
which turns around the tree (see Figure \ref{turnaround}). This is the key point.
This trace of an operator of norm smaller than 1 is bounded by $N$.

It remains now to collect the other factors. There is an $N$ factor for each vertex of the tree
and a $\vert -i \sqrt{\lambda /N} \vert $ factor for each half line of the tree. 
Collecting all the $N$ factors we get therefore a a $N^{2}$ global, $n$ independent 
factor as should be the case
for vacuum graphs in this matrix $\Phi^4$ theory, times $\lambda^{n-1}$.
\qed

We can now integrate the previous bound over the complicated measure $d\nu_T$
and over the $\{w_\ell\}$ parameters. 
But since our bound is independent of ${\sigma^v}$ and $w$'s, since the measure $d\nu(\sigma)$ 
is normalized, and each $w_\ell$ integral runs from 0 to 1, the bound goes through.

Finally by Cayley's theorem the sum over trees costs $n! \prod_v \frac{1}{(k_v -1)!} $
The $n!$ cancels with the $1/n!$ and we remain with a geometric series bounded by 
$N^{2}  \sum_{n \ge 1}\lambda^{n-1}$ hence convergent for $\lambda<1$.

This completes the proof of Theorem \ref{maintheor}. \qed

\section{Uniform Borel summability} 

Rotating to complex $\lambda$ and Taylor expanding out a fixed number of vertices proves 
Borel summability in $\lambda$ \emph{uniformly in} $N$. 

\medskip
\noindent{\bf Definition}
\medskip
{\it
A family $f_N$ of functions is called Borel summable in $\lambda$ {\bf uniformly in $N$} if 

\begin{itemize}
\item
Each $f_N$ is analytic in an $N$ independent disk 
$D_R = \{ \lambda \vert {\rm Re}\, \lambda^{-1} > 1/R\}$;

\item Each $f_N$ admits an asymptotic power series $\sum_k a_{N,k}\lambda^k  $
(its Taylor series at the origin) hence:
\begin{equation}  f_N(\lambda) = \sum_{k=0}^{r-1} a_{N,k} \lambda^k + R_{N,r} (\lambda)  
\end{equation} 
such that the bound
\begin{equation} \label{taylorremainder}  \vert R_{N,r} (\lambda) \vert \le A_N \sigma^r r! \vert \lambda \vert^r 
\end{equation} 
holds uniformly in $r$ and $\lambda \in D_R$, for some constant $\sigma \ge 0$ independent
of $N$ and constants $A_N \ge 0$ which may depend on $N$.

\end{itemize}
}

Then every $f_N$ is Borel summable \cite{Sok}, i.e. the power series 
$\sum_k a_{N,k} {t^k \over k!}$ converges for $\vert t \vert < {1 \over \sigma}$. It
defines a function $B_N(t)$ which has an analytic continuation in the $N$ independent strip 
$S_{\sigma} = \{t \vert {\rm \ dist \ } (t, {{\mathbb R}}^+) < {1 \over \sigma}\}$.
Each such function satisfies the bound
\begin{equation} \vert B_N(t)  \vert \le { \rm B_N} e^{t \over R} \quad {\rm for \ } 
t \in { {\mathbb R}}^+  
\end{equation}
for some constants $B_N \ge 0$ which may depend on $N$.
Finally each $f_N$ is represented by the following absolutely convergent integral:
\begin{equation}  f_N(\lambda) = {1 \over \lambda} \int_{0}^{\infty} e^{-{t \over \lambda}}  B_N(t) dt \quad\quad
 \quad {\rm for \ } \lambda\in C_R .
\end{equation}

\begin{theorem}\label{borelunif}
The series for $Z(\lambda, N)$ is uniformly Borel summable with respect
to the slice index $N$.
\end{theorem}

\noindent{\bf Proof}
It is easy to obtain uniform analyticity for $\Re \lambda >0$ and $\vert \lambda\vert $
small enough, a region which obviously contains a disk $D_R$.
Indeed all one has to do is to reproduce the previous argument but 
adding that for $H$ Hermitian, the operator
$(1+i e^{i \theta} H)^{-1}$ is bounded by 2 for $\vert \theta \vert \le \pi /4$.
Indeed if $\pi/4 \le {\rm Arg} z \le 3\pi/4 $, we have $\vert (1+i z )^{-1}\vert \le \sqrt{2}$.

Then the uniform bounds (\ref{taylorremainder}) follow from 
expanding the product of resolvents in (\ref{loopvertex}) up to order $r-2(n-1)$
in $\lambda$. by an explicit Taylor formula with integral remainder followed 
by explicit Wick contractions. The sum over the contractions leads to the 
$\sigma^r r!$ factor in (\ref{taylorremainder}); in our case the constants $A_N = K. N^2$
actually depend on $N$ but this is allowed by our definition of uniform Borel summability.
\qed

\section{Correlation Functions}

To obtain the connected functions with
external legs we need to add resolvents to the initial loop vertices.  A resolvent is
an operator $C(\sigma^r, m_1, m_2) $, which can depend on only two indices
because in a matrix model every entering index must go out. 
The connected functions $S^c(m_1, ..., m_{2p}) $ therefore depend only on
$2m$, not $4m$ indices. They
are obtained from the normalized functions by the standard procedure. We 
have the analog of formula \ref{treeformul} for these connected functions:
\begin{theorem}

\begin{eqnarray}\label{treeformulext}
&&S^{c}(m_1, ..., m_{2p}) 
=\sum_{\pi} \sum_{n=1}^{\infty} \sum_T \bigg\{ \prod_{\ell\in T}   
\big[ \int_0^1 dw_\ell 
\sum_{i_\ell, j_\ell, k_\ell, l_\ell} \big]\bigg\} 
\nonumber 
\int  d\nu_T (\{\sigma^v\}, \{ w \}) 
 \\&&
 \bigg\{ \prod_{\ell\in T} \big[ \delta _{i_\ell l_\ell} \delta_{j_\ell k_\ell}
 \frac{\delta}{\delta \sigma^{v(\ell)}_{i_\ell, j_\ell}}
 \frac{\delta}{\delta \sigma^{v'(\ell)}_{k_\ell, l_\ell}} 
\big] \bigg\}
 \bigg\{\prod_v V_v \prod_{r=1}^{p}   C_{j}(\sigma_{r}, z_{\pi(r,1)}, z_{\pi(r,2)})\bigg\} 
\end{eqnarray}
where 
\begin{itemize}
\item the sum over $\pi $ runs over the pairings of the $2p$ external variables
into pairs $(z_{\pi(r,1)}, z_{\pi(r,2)})$, $r=1,...,p$, 

\item each line $\ell$ of the tree joins two different loop vertices or resolvents 
$V_{v(\ell)}$ and $V_{v'(\ell)}$,

\item the sum is over trees joining the $n+p$ loop vertices and resolvents, which have therefore
$n+p-1$ lines,

\item the measure $d\nu_T (\{\sigma^v\}, \{\sigma_r\}, \{ w \})  $ over the vector fields 
$\{\sigma^\alpha\}$ has covariance 
$$<\sigma^\alpha_{ij} \sigma^{\alpha'}_{kl}>= \delta _{il} \delta_{jk}
w^T (\alpha, \alpha', \{ w\})$$
where again for $\alpha, \alpha'\in \{v\}, \{r\}$, $w^T (\alpha, \alpha', \{ w\})$ is 1 if $\alpha=\alpha'$,
and the infimum of the $w_\ell$ for $\ell$ running over the unique path from 
$\alpha$ to $\alpha'$ in $T$
if $\alpha\ne \alpha'$.

\end{itemize}
\end{theorem}

This expansion is convergent exactly as the initial one and we get:

\begin{theorem}
The series (\ref{treeformulext}) is absolutely convergent for $\lambda$ small enough, and we have:
\begin{equation}
\vert  S^{c}(m_1, ..., m_{2p}) \vert \le K \,(2p)!! N^{2-p} .
\end{equation}
\end{theorem}

\section{Further topics}

\subsection{Symmetric or Hermitian matrix models}

Interacting GOE and  GUE models can be treated along the same lines. Let us consider for instance
the same model than (\ref{functional}) but with $\Phi=\Phi^\star$ now an Hermitian matrix.
We have no longer a canonical distinction between left and right indices so that 
the intermediate field operator acts on both sides, but it is still anti-Hermitian.
The vertex operator (\ref{vertex}) 
is therefore replaced by
\begin{equation}
V= -\Tr \log (1\otimes 1 + \frac{i}{2} \sqrt{\frac{\lambda}{N}} [ \sigma \otimes 1 + 1 \otimes \sigma]),
\end{equation}
so that each loop vertex is no longer simply proportional to $N$
because of e.g. the left trace. But any tree is planar so one can still draw 
the tree between loop vertices on a plane, as in Figure 3.
The total number of traces of products of $(1 +i H)^{-1}$ operators for a tree on $n$ vertices
still remains $n+1$ by Euler formula. Indeed Euler formula says 
$2-2g = V-L+F$, where $g$ is the genus and $F$ is the number of faces, each costing $N$.
But graphs of genus 0 as those of Figure 3 contain 
$2(n-1)$ vertices (of the cubic type), and two kinds of lines,
the $n-1$ lines of the tree and the $\sum_v k_v = 2(n-1)$ \emph{resolvent lines}.
Therefore $F = 2 -2(n-1) + (n-1+2(n-1)) = n+1 $  so that all the results
of the previous sections remain valid. 

\begin{figure}[!htb]\label{inouttree}
\centering
\includegraphics[scale=0.3,angle=-90]{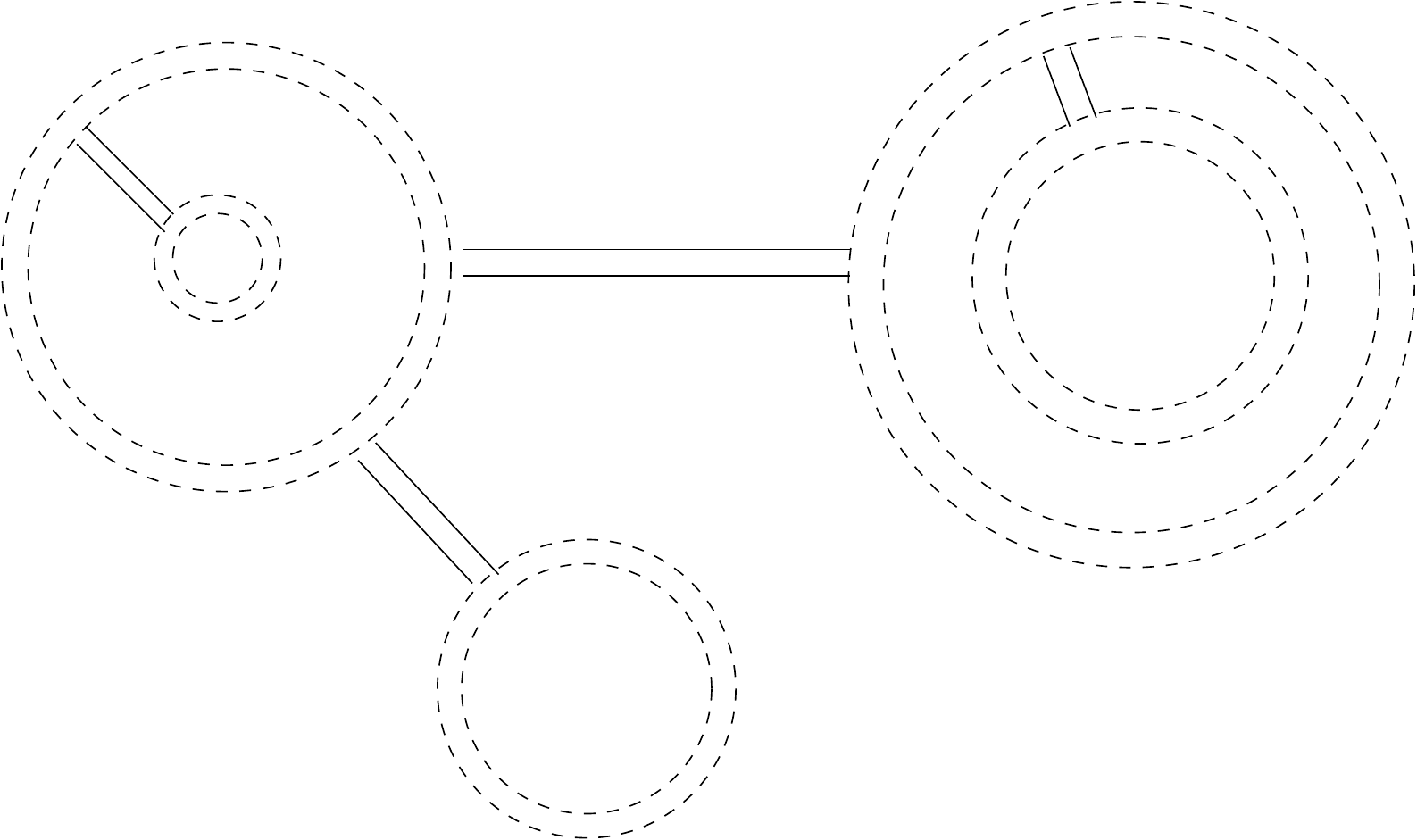}
\caption{A tree with five loop vertices joined by four tree lines, eight cubic vertices, and eight resolvent lines
which is a contribution in the Hermitian case.}
\end{figure}

\subsection{Genus expansion}

The genus expansion which lies at the root of matrix models can be generalized constructively.
We can indeed expand the resolvents on the external loop which turns around the tree
in Figures 1 or 3, and Wick-contract one at a time the $\sigma$ fields produced.
If we were to perform this to all orders the expansion would diverge.
However we can also contract until a fixed number of non-planar crossings are generated,
and then stop. We call this expansion a ``rosette expansion".
It does not diverge and allows to extract 
the $1/N$ expansion up to a fixed genus $g$, plus  a smaller remainder. 
For instance for the normalization one obtains a representation
\begin{equation}
Z(\lambda, N)= N^2 \bigg[  \sum_{k=0}^{g} N^{-2k} a_{k,\lambda}  + R_{g+1}(\lambda, N )
\bigg]
\end{equation}
where $a_k(\lambda)$, the sum over all vacuum graphs of genus $k$, is an analytic
function of $\lambda$  and $R_{g+1}(\lambda, N )$ 
is a convergent series whose sum is bounded by $O(N^{-2(g+1)})$ and is again
Borel summable in $\lambda$
uniformly in $N$.

This remark is essential to construct $\phi^{\star 4}_4$ through this method. 
We need indeed to identify the planar contributions with a single broken face
and two or four external legs because they are 
the only ones which need to be renormalized, and also the only ones which 
can be renormalized (because only planar graphs with a single broken face
look like Moyal products
when seen from lower renormalization group scales \cite{Riv2}). 
It is therefore essential to have a method which can extract them from the rest of the expansion
without violating the constructive golden rule.
This can be done through the rosette expansion sketched above.

\subsection{Decay of correlations in quasi-matrix models}

To fully construct $\phi^{\star 4}_4$  we have to take into account
the fact that the propagator of $\phi^{\star 4}_4$ in the matrix base
does not exactly conserve matrix indices \cite{GrWu2}, except at $\Omega=1$, where
$\Omega$ is the Grosse-Wulkenhaar parameter. 

It is therefore essential to show not only uniform convergence 
but also decay of connected functions with respect to external
matrix indices in this kind of models. This should not be too difficult using
iterated resolvents bounds, as is shown in \cite{MR} in the  case of 
ordinary $\phi^4$ on commutative space.

\subsection{Multiscale Analysis}

To fully construct $\phi^{\star 4}_4$  we have also to generalize the single $N$ 
analysis of this paper to a multiscale analysis such as the one of \cite{RVW}.
This requires to optimize as usually the tree expansion over all the scales
so that connected functions of higher scales are always correctly connected through the tree. 

In fact the $\phi^{\star 4}_4$  can presumably also be built as easily in $x$ space representation
by a slight modification of the matrix 
argument. Indeed a Moyal $\phi^4$ vertex can be decomposed in terms
of an intermediate ultralocal real field with a $\Tr \bar \phi \star \phi \star  \sigma$ interaction. 
This can again be done in two ways by duality. The new vertex is anti-hermitian again 
as a kernel between the $\bar \phi$ and $\phi $ points. The bosonic covariance 
of the $\phi$ field is a Mehler kernel that can be
easily broken in square roots. We obtain loops of Mehler kernels sandwiched 
between operators of the $(1+iH)^{-1}$ type. We expect therefore 
all constructive aspects to be also doable in $x$-space \cite{MR2}.

Since our loop vertex expansion seems very well suited to treat both 
large $N$ vector and large $N$ matrix
limits, we expect that it is the right tool to glue different regimes of the renormalization group
governed respectively e.g. in the ultraviolet regime by a small coupling expansion and 
in the infrared by a ``non-perturbative" large $N$ expansion of vector or matrix type. This gluing problem 
occurs for the vector case in many different physical contexts, from mass generation of the two-dimensional 
Gross-Neveu \cite{KMR} or non-linear $\sigma$-model \cite{K} to the BCS theory of supraconductivity \cite{FMRT}. 
Confinement itself could be a matrix version of the same gluing problem \cite{Hoo}. 
All such gluing problems have been considered until now too complicated in practice
for a rigorous (i.e. constructive) analysis. We hope that this might change over the coming years.

\medskip
\noindent{\bf Acknowledgments} We thank A. Abdesselam, M. Disertori, R. Gurau, J. Magnen
and F. Vignes-Tourneret for many useful discussions which lead to the slow maturation 
of this paper.

\end{document}